\documentclass[twocolumn,showpacs,preprintnumbers,amsmath,amssymb]{revtex4}

\usepackage{graphicx}
\usepackage{dcolumn}
\usepackage{bm}

\begin{document}

\title{Generating THz light from  microwave}

\author{S. Son}
\affiliation{169 Snowden Lane, Princeton, NJ 08540}

\begin{abstract}
A new way of the THz light generation is considered.
 A  tera-hertz light is emitted from the interaction between a high frequency microwave  and a relativistic electron beam, via the backward Raman scattering. 
The up-shifting of the frequency occurs through the relativistic Doppler's effect.  This scheme may lead to a cheap and compact tera-hertz light sources.
\end{abstract}

\pacs{42.55.Vc, 42.65.Ky, 52.38.-r, 52.35.Hr}       

\maketitle

\section{Introduction} 

There have been  exploding 
 interests for a  commercially-viable  THz light source, as it would create new industries in the fusion plasma diagnostic, the molecular spectroscopy, the tele-communication and many others~\cite{siegel,siegel2,  siegel3, booske,radar, diagnostic, security}.
Numerous THz light sources have been invented so far~\cite{Tilborg,Zheng,Reimann, gyrotron, gyrotron2, gyrotron3,magnetron, qlaser, qlaser3, freelaser, freelaser2, colson, songamma,Gallardo,sonttera}, including the free electron laser~\cite{ freelaser, freelaser2, colson}, the gyrotron, the vacuum electronic device~\cite{ gyrotron, gyrotron2, gyrotron3,magnetron}, the quantum cascade laser~\cite{ qlaser, qlaser3} and the laser-based technology. 
Surprisingly, however, 
THz light sources 
are often neither practical nor intense  enough. 
The inability in producing THz light intense enough   
is referred to as the ``THz Gap''~\cite{booske}.
In addition, current light sources often require  expensive strong magnets and accelerators, or often need to be operated in extremely low-temperature.
Significant progresses, in enhancing the intensity (power), reducing the cost (size) or increasing the efficiency,  are necessary for its commerical applications.
 
In this paper, the author proposes a new way of generating  THz light. 
The author considers a situation where
1) an high frequency microwave  \textit{counter-travels} with a relativistic electron beam, 2) the microwave is reflected  by the electron beam via a \textbf{certain interaction}. 
and 3) the reflected light is in the THz frequency   due to the Doppler's frequency up-shift. 
As for the \textbf{certain interaction} in 2), 
two scenarios are considered. 
The first one is the Raman scattering;  a)  Langmuir waves are excited inside the electron beam via two-stream instability or pondermotive potential by the visible light lasers and b) the microwave is reflected  via the Raman scattering of  the excited Langmuir waves. 
The second one is the total reflection; As the electron beam density reaches the critical density of the microwave, the microwave will be reflected as in the case that the visible light laser is reflected against the metal. 
The analysis in this paper suggests that
  the proposed  THz light source could be  small, cheap and efficient and that
the light intense might be as high as any exitent light source.  

\section{THz light from a high-freqeuncy microwave  via the Dopper's effect}

Considering  a microwave 
with the frequency $\omega_{m0} $ (wave vector $k_{m0}$) and  a  counter-propagating electron beam with the electron density   $n_0$  and the relativistic factor $\gamma_0= (1-v_0^2/c^2)^{-1/2}$,  where $v_0$ is the velocity of the electron beam, 
the electron beam density decreases to  
$n_1 = n_0 / \gamma_0$ 
in the co-moving frame  due to the Doppler's effect.
The microwave in the co-moving frame satisfies
the usual dispersion relationship, $\omega_1^2 =   \omega_{pe}^2/\gamma_0+ c^2 k_1^2$, where 
$\omega_1$ ($k_1$) is the waves frequency (vector) and $\omega_{pe}^2 = 4 \pi n_e e^2 / m_e $ is the Langmuir wave frequency with $n_e$ ($T_e$) being the electron density (temperature) in the laboratory frame.

 Denoting the wave vector  (the corresponding wave frequency) of 
the microwave (the reflected wave) in the co-moving frame 
as  $k_{m1}$, $k_{T1} $, $\omega_{m1} $ and   $\omega_{T1} $, 
and those in the laboratory-frame counterparts  
 as  $k_{m0}$, $k_{T0} $,  $\omega_{m0} $ and $\omega_{T0} $,
the Lorentz transform prescribes the following relationship: 
\begin{eqnarray} 
\omega_{m0} &=& \gamma_0 \left[ \sqrt{\omega_{pe}^2/\gamma_0 + c^2 k_{m1}^2 } - vk_{m1} \right] \mathrm{,}  \label{eq:lorentz1} \\  \nonumber \\
k_{m0} &=&  \gamma_0 \left[ k_{m1} - \frac{\omega_{m1} }{c}  \frac{v_0}{c} \right] \mathrm{,} \label{eq:lorentz2} \\ \nonumber \\ 
 \omega_{T0} &=& \gamma_0 \left[ \sqrt{\omega_{pe}^2/\gamma_0 + c^2 k_{T1}^2 } + 
vk_{T1} \right] \mathrm{,}  \label{eq:lorentz3} \\  \nonumber \\ 
k_{T0} &=&  \gamma_0 \left[ k_{T1} + \frac{\omega_{T1} }{c}  \frac{v_0}{c} \right]\mathrm{.} \label{eq:lorentz4} \\ \nonumber 
\end{eqnarray}
The only difference between the microwave and the reflected wave is the sign because the direction of the propagation is opposite. 
Using equations (\ref{eq:lorentz1}) to (\ref{eq:lorentz4}), 
the microwave (the reflected wave) can be transformed between the co-moving and the laboratory frames.

The energy and the momentum conservation of the reflection physics is represented as   
\begin{eqnarray} 
 \omega_{m1} &=& \omega_{T1} + \omega_{3}  \nonumber 
\mathrm{,} \nonumber \\  
k_{m1} &=& k_{T1} +k_3\mathrm{,} \label{eq:cons}  
\end{eqnarray}
where $k_3$ ( $\omega_{3}$ )is the plasma wave vector (frequency).
In the case of the plasma wave interaction, $\omega_{3}$ is the plasma wave frequency.  
In the case of the total reflection, $\omega_{3}=0$ and $k_3 = -2 k_{m1} $  because the total reflection changes not  the frequency but the propagation direction. 
For a given microwave frequency $\omega_{m0} $, 
$k_{m1} $ ($\omega_{m1} $) is  obtained from  Eqs. (\ref{eq:lorentz1}) and (\ref{eq:lorentz2}), 
  $k_{T1} $ ($\omega_{T1}$) is  from  equation (\ref{eq:cons}) and    
$k_{T0} $ ($\omega_{T0}$) is  from 
 equation (\ref{eq:lorentz3}) and (\ref{eq:lorentz4}). 
In either case of  the plasma wave interaction or the total reflection,
it can be derived with assumption of $\omega_{3} \ll \omega_{m0} $: 
\begin{equation} 
\omega_{T0} \cong \left(\frac{1+v_0/c}{1-v_0/c} \right) \left( \omega_{m0} - \frac{ \omega_{3}}{2 \gamma_0}\right) \mathrm{,}\label{eq:down1}
\end{equation}   
Equation~(\ref{eq:down1}) describes the frequency up-shifting of the microwave
into the  THz regime, via the relativistic Doppler effect. For an example, 
for an 100 keV electron beam,  the frequency up-shift is a factor of 3.  
If the microwave has a frequency of 300 GHz ($\omega_{m0}$),  the reflected light has the 0.9 THz frequency ($\omega_{T0}$). 

\section{Reflection via a plasma wave}

The first  reflection method   is the backward Raman scattering. 
There are various methods to excite a plasma wave, which include the two-stream instability or the beating of  intense lasers. 
Once a plasma wave is excited, the reflection occurs via the interaction between the plasma wave and the microwave. 
An example of the two-stream instability  will be discussed. However, it will be appreciated that 
the two-stream instability is just an illustration for  exciting  plasma waves.

 The wave vector of the microwave in the co-moving frame with the electron beam,   from Eq.~(\ref{eq:lorentz2}), is  
\begin{equation}
k_{m1} = \frac{k_{m0}}{ \gamma_0 (1-v_0/c) }\mathrm{.} \label{eq:microl}
\end{equation} 
The plasma wave vector for the reflection is 
\begin{equation}
k_3 \cong 2 k_{m1} =  2  \frac{k_{m0}}{ \gamma_0 (1-v_0/c) }\mathrm{.} \label{eq:microl2}
\end{equation}
One preferred condition for such a plasma wave to be existent in the beam is that the Debye wave vector is higher than the one of the plasma wave:
\begin{equation} 
 k_3 \lambda_{de}  \le 1 \label{eq:lan}\mathrm{,}
\end{equation} 
where $\lambda_{de}^2 =4\pi n_1 e^2 / T_e  =  4\pi n_0 e^2 / T_e \gamma_0 $ is the Debye length and $T_e $ is the electron temperature from the electron beam spread. 
Since  $\lambda_{de}  \cong 7.6 \times 10^2  n_e^{-1/2}\sqrt{T_e} $ where 
   the electron density in the unit of cc and the $T_e$ in the unit of the electron volt,  the condition Eq.~(\ref{eq:lan}) can be re-casted as 

\begin{equation} 3.2 \times 10^{4} n_e^{-2/1} \sqrt{T_e} \frac{f_{100}}{\gamma_0 (1-v_0/c) } < 1 \mathrm{.} \label{eq:lan2}
\end{equation}  
where $f_{100}$ is the frequency of the microwave in the laboratory frame normalized by 100 GHz. For an example,  for  a 100 keV electron beam, $n_e > 2 \times 10^{10}  T_e$ per cc.

 In exciting a Langmuir wave via the two-stream instability,
multiple electron beams with different drifts would be ideal. 
A Langmuir wave given in Eq.~(\ref{eq:microl2})
is hard to excite in a rare-dense electron beam as its wave vector might be too
 high for the electron beam to support as a collective  wave.  
Therefore, it is preferred that the Langmuir waves with high wave vector
 is excited in a less dense plasma.

Estimation of  the highest wave vector of the Langmuir waves unstable to the two-stream instability is estimated in the following. 
The criteria of the two-stream instability would be that 1) there is a local maxima of the dielectric function $\epsilon$ as a function of $\omega$ for a fixed $k$ and 2) the value of the local maxima is less than zero. 
The longitudinal dielectric function of the  beams  is   

\begin{equation} 
\epsilon(\mathbf{k}, \omega) = 1 + \frac{4 \pi e^2 }{k^2} \Sigma \chi_i \mathrm{.}
\end{equation} 
where the summation is over the group of particle species and $  \chi_i $ is the particle susceptibility. In classical plasmas, the susceptibility is given as 

\begin{equation}
 \chi_i^C(k, \omega) = \frac{n_iZ_i^2}{m_i} \int \left[ \frac{ \mathbf{k} \cdot \mathbf{\nabla}_v f_i }{\omega - \mathbf{k} \cdot \mathbf{v} }\right]d^3 \mathbf{v} 
\end{equation} 
where $m_i$ ($Z_i$, $n_i$) is the particle mass (charge, density) and $f_i $ is the distribution with the normalization $\int f_i d^3 \mathbf{v} = 1$. For the case of our two group of electrons, it is given as 
\begin{equation} 
\epsilon =  1 + (4 \pi e^2/k^2) ( \chi_e^C(\omega, k) +  \chi_e^C(\omega-\mathbf{k}\cdot \mathbf{\delta v}, k) \mathrm{.}
 \label{eq:ele}
\end{equation} 
where $\mathbf{\delta v}$  is the difference in the drift velocity between two electron beams in the co-moving frame.

The stability analysis  for the various beam density and temperature has been conducted based on Eq.~(\ref{eq:ele}). For its convenience, it is assumed that two electron beams have the same 
electron density, but it will be appreciated that the two beams does not have to have the same electron density.
From the analysis conducted,  the optimal drift velocity $\delta v $ can be estimated as 
\begin{equation}
 2 < \sqrt{\gamma_0 }k_{3} \delta v/\omega_{pe} < 10 \mathrm{.} \label{eq:opt}
\end{equation} 
Eq.~(\ref{eq:opt}) in conjunction with Eq.~(\ref{eq:lan2}) provides the estimate for the optimal parameter of exciting the Langmuir wave for the purpose of the THz generation. 

As the first example, consider a 100 keV electron beam and the microwave with  300 GHz.   
If $T_e \cong 10  \ \mathrm{eV} $, 
 Eq.~(\ref{eq:lan2}) is $n_e > 10^{12}  / \mathrm{cc} $.  If the drift velocity between the electron beams corresponds to 5 keV, the Langmuir wave appropriate for the reflection is unstable. The emitted light has the frequency of 0.9 THz. 

As the second example, consider a 100 keV electron beam and the microwave with 300 GHz.  
If $T_e \cong 1  \ \mathrm{keV} $, 
 Eq.~(\ref{eq:lan2}) is  $n_e > 5  10^{14}  / \mathrm{cc} $.
 If the drift velocity between the electron beams corresponds to 5 keV, the Langmuir wave appropriate for the reflection  is unstable. The emitted light has the frequency of  0.9 THz. 

As the third example, consider a 500 keV electron beam and the microwave with  50 GHz.  
If $T_e \cong 1  \ \mathrm{keV} $, 
Eq.~(\ref{eq:lan2}) is $n_e > 2 \times   10^{12}  / \mathrm{cc} $.
 If the drift velocity between the electron beams corresponds to  a 5 keV, the Langmuir wave appropriate for the reflection is unstable. The emitted light has the frequency of 0.8 THz. 

 Once the Langmuir wave is excited, the  channeling the microwave into the THz light can be analyzed using the fluid equation. 
The 1-D BRS three-wave interaction 
in the \textit{co-moving} frame between the microwave,  
the THz light  and a Langmuir wave  
is described  by~\cite{McKinstrie}:
\begin{eqnarray}
\left( \frac{\partial }{\partial t} + v_m \frac{\partial}{\partial x} + \nu_1\right)A_m  = -ic_m A_T A_3  \nonumber \mathrm{,}\\
\left( \frac{\partial }{\partial t} + v_T \frac{\partial}{\partial x} + \nu_2\right)A_T  = -ic_T A_m A^*_3   \label{eq:2} \mathrm{,} \\
\left( \frac{\partial }{\partial t} + v_3 \frac{\partial}{\partial x} + \nu_3\right)A_3  = -ic_3 A_m A^*_T  
\nonumber \mathrm{,}
\end{eqnarray}
where $A_i= eE_{i1}/m_e\omega_{i1}c$  is 
the ratio of  the electron quiver velocity of the microwave ($i=m$)
and the THz light ($i=T$)  relative to the velocity of the light $c$,
 $E_{i1}$ is the electric field of the E\&M pulse, 
 $A_3 = \delta n_1/n_1$ is the the Langmuir wave amplitude,
$\nu_1 $ ($\nu_2$) is the rate of the inverse bremsstrahlung  
of the microwave (THz light), 
$\nu_3$ is the Langmuir wave  decay rate, 
$ c_i = \omega_3^2/ 2 \omega_{i1}$ for $i=m, T$, $c_3 = (ck_3)^2/2\omega_3$, 
$\omega_{m}$ ($\omega_{T}$) is the wave frequency of the microwave (the THz light) and  $\omega_{3} \cong \omega_{pe} / \sqrt{\gamma_0} $  ($k_3$) is  the plasmon  wave frequency (vector). 

The first one  in  Eq.~(\ref{eq:2}) is the  most relevant 
 as the excited Langmuir wave is given by $A_3= \delta n_e /n_e$.  
 The mean-free path of the microwave to the BRS is estimated from Eq.~(\ref{eq:2}) to be 
\begin{equation} 
l_b \cong c (2 \omega_{T1}  /\omega_3^2 ) (1/A_3) 
\mathrm{.} \label{eq:mean}
\end{equation}
The mean-free-path from 
the Thomson scattering (the Compton scattering) is  $ l_t \cong 1/n\sigma_t $ with $\sigma_t = (mc^2 / e^2)^2 $.  For an example, when $n_1 \cong 5\times 10^{11} / \mathrm{cc} $,   
 $l_t \cong 0.1 \times 10^{12} / n_{12} \ \mathrm{cm} $ and $l_b \cong 100 \times F_{T} / (A_3 n_{12} ) \ \mathrm{cm}$, 
where $F_{T}$  is the THz light (the microwave) frequency normalized by 1 THz and $n_{12} $ is the beam electron density normalized by $10^{12}$ per cc.  Even for $A_3 \cong 0.001 $, the THz   radiation by the BRS is million times stronger than the Thomson scattering or $l_t \gg l_b $. 

Up to recently, the most of the experiments or simulations observed the intensity of  $A_3= \delta n_e /n_e > 0.1$ often. The theoretically possible peak intensity might be an interesting research topic.  If  $A_3\cong 0.1$ and $n_{12} = 1$, then $l_b < 10^{3} F_{T}  \ (\mathrm{cm})$. In the co-moving frame, 
$F_{T}<1$ as it is the frequency of the microwave in the moving frame.   
The ratio between the interaction region $L$ and the mean-free-path $l_b$ is roughly the percentage of the microwave converted into the THz light; 

\begin{equation} 
\eta =  \frac{L}{l_b} \mathrm{.}
\end{equation}
If $ L \cong 10 \ \mathrm{cm}$ and $A_3 \cong 0.1$, $\eta \cong 0.01$ so that a few percentages of the microwave is converted into the THz light. 
 If $n_{12} = 100$ and $A_3\cong  0.1$,  then $l_b < 10 F_{T}  \ (\mathrm{cm})$. If  $L \cong 10 \ \mathrm{cm}$, then the most energy  of the microwave is converted into the THz light ($\eta  \cong 1$).

As shown in the above examples,  the preferred electron beam density is moderate and the electron beam energy does not need to be very high, comparable to 100 keV. 
Such beam can be generated  economically  from a compact source such as   a conventional cathode and a virtual cathode. The conventional accelerator or the laser-metal interaction can be also used.

Based on the above analysis and examples, 
plausible input microwave (output THz) energy can be estimated as follows.
For an electron beam with the electron density $n_{12}$ and the electron energy 100 keV, the energy density of the electron beam is $I_b = (100 \ \mathrm{keV}) \times (n_e) \times c  \cong 100 \times n_{12}\ (\mathrm{MW} / \mathrm{cm}^2$), where $I_b$ is given in the unit of watts per square centimeter.  If the cross-section of the electron beam is $A$, the electron beam power is given as 
$P_b \cong   100 \ A n_{12}  \ (\mathrm{MW}$).  Denoting the beam duration as $\tau$, the total electron beam energy is 
 $E_b \cong 100 \ A n_{12} \tau \ (\mathrm{MJ}) $.

For its convenience, it will be assumed that the duration of the microwave and cross-section of the microwave is comparable to those of the electron beam. Then,  
the intensity of the microwave is given as $I_m = I_{6} \ (\mathrm{MW} / \mathrm{cm}^2)$, where $I_{6}$ is normalized by $\mathrm{MW} / \mathrm{cm}^2$. 

 If the electron beam and microwave is injected continously, 
the output power of the THz light can be estimate as  

\begin{equation} 
P_T = R I_m A \eta = R I_6 A \eta  \ \left(\mathrm{MW} / \mathrm{cm}^2\right) \mathrm{,}
\end{equation}
where $R$ is the Doppler's frequency up-shift factor, $\eta \cong L/l_b$ is the conversion efficiency of the microwave into the a THz light and  
$A$ is the cross-section. For an example, 
if $A=0.1 \ (\mathrm{cm}^2)$, $I_6 =1 $,  $L \cong 10 \ (\mathrm{cm})$ and $\eta = 0.01$,  the electron beam power is 10 MW and the THz output power is 1 KW. 

 Finally, 
 the efficiency of the THz output  power to the input power is given as 
\begin{equation} 
 Eff = P_T/(P_b + P_m)  \cong 10^{-2}\times R \eta I_{6} / n_{12} \label{eq:eff}  \mathrm{.}
\end{equation}
It is noted that the beam duration and the cross-section cancel out for the efficiency ratio, as the ouput and the input energy are all proportional to them. 
If $0.01<\eta<1$,  $10^{-4} \times  R I_{6} < Eff < 10^{-2}\times  R I_{6} $ for $n_{12} =1 $ and 
  $10^{-6} \times R I_{6} < Eff < 10^{-4} \times R I_{6} $ for $n_{12} =100 $. 
The efficiency  $Eff$ is proportional to $\eta $ but inversely proportional to the electron density. 
With the electron density increasing, 
$\eta \cong L/l_b$ ($n_{12}$) decreases (increases) as the mean-free-path $l_b$  decreases in higher electron beam density.

 In continous operation of an 1 MW electron beam, the output power of the THz light can achieve  between 10 kW and 100 W, if $1< I_{6}<100$ and $\eta \cong 0.01$. This power is much higher than the current technological limit that is  $10^{-3}$W.  
 However, in continous operation, the electron beam power $P_b$ is very high  for an compact THz light.  
Instead, it might be desirable to decrease the cross-section and operate in a pulsed operation. 
If the pulsed duration is 1 nanosecond and $A=0.1 \ \mathrm{cm}^2$, 
the electron beam energy per shot is $ 0.01 \  \mathrm{J}$ and the THz energy per shot is  $10^{-4} \times \eta I_{6}/n_{12} \ \mathrm{J}$.
If the duration is a microsecond with $A=0.1 \ \mathrm{cm}^2$, 
the electron beam energy per shot is 10 joule and the THz energy per shot is  $0.1 \times \eta I_{6}/n_{12} \ \mathrm{J}$.
If $\eta \cong 0.01$ and $I_{6} =1 $, the output THz energy per shot can reach $10^{-3}$ ($10^{-6}$) joule from an electron beam with the energy of 10 (0.01) joules.  If the electron beam has $ 10^{3} \  \mathrm{J}$, the THz light output energy could be comparable to $ 1 \  \mathrm{J}$.

 As given in Eq.~(\ref{eq:eff}), the most key physical parameter in the efficiency ratio is 
the microwave intensity $I_{6}$.  The more intense the microwave is, the more efficient the THz light source is.  If $I_{6} = 0.001$, the output THz energy is $10 \  \mathrm{\mu J}$ from from an electron beam with the energy of $10 \ \mathrm{J}$. The THz light with $10 \  \mathrm{\mu J}$ per shot  is still very high in comparison to any current powerful light source. However,  
if  $I_{6} = 100$,  the output THz energy is $0.1 \  \mathrm{J}$ from an electron beam with $10 \  \mathrm{J}$, which is larger than any THz light source by a few order of magnitude. Therefore, it is better to have the microwave as intense as possible.  
The microwave intensity can easily reach $I_6 = 1$ in continous operation~\cite{hidaka, kumar}.
However, in order to operate 100 GHz gyrotron in a continous fashion with  $I_6 \cong 1$,
the gyrotron needs to be cooled cryogenically, which is challenging. 
There could be various methods to increase the intensity $I_{6}$ without 
cryogenic cooling. Firstly, by doing the pulsed operation, 
the temporal peak intensity could be increase dramatically. Secondly, by capturing the microwave energy in a resonant cavity, the intensity could be incrase as much as Q factor.  Thirdly, the combination of the pulsed operation and the utilization of the resonant cavity would increase the intensity further.
If  $I_{6} = 100$, the efficiency ratio could reach almost 0.01 if $\eta \cong 0.01$

\section{Total Reflection}

The second reflection method considered  is the total reflection by an electron beam.
If the electron density of an electron beam is higher than the critical density,
a microwave, encontering an electron beam in a opposite direction,  is reflected  as a visible light laser getting reflected  in the metal surface.
The critical density for the reflection of the microwave is given as $\omega_{pe}/\gamma_0 > \omega_{m1} $ or 

\begin{equation}
n_{e} > n_c =  1.2 \times 10^{14}\left(\frac{f_{100}^2}{\gamma_0}\right)\left( \frac{1+v_0/c}{1-v_0/c}\right) \mathrm{,} \label{eq:critical} 
\end{equation}
where $f_{100}$ is the frequency of the microwave normalized by 100 GHz. 

If it is for a visible light laser or a infra-red laser, 
$300 \ge f_{100} \ge 30 $ so that $    10^{19} \ge   n_{c} \ge 10^{17} $ per cc. However,  the microwave has $f_{100} \cong 1 $ so that $n_c \cong 10^{14}$ per cc, which is much less dense and can be created with much ease.  
One of the important idea in this paper is 
to reduce the preferred electron beam density by utilizing a high-frequency microwave instead of a visible light laser. 

There have been much progresses in generating highly collimated dense electron beam via the interaction between the metal layers and an intense laser~\cite{monoelectron, ebeam}. 
Those mono-energetic electron beams are preferred for the purpose of the fast ignition in the inertial confinement fusion~\cite{tabak}. 
As an intense laser encounters a plasma or metal, electrons are blown out.  
Those electrons are blown out in the back of the metals or the front of the metals and might be collimated or not. 
Under current technologies, 
 the electron beam reaches the energy between a few hundred keV and a few MeV and the density as high as $10^{20}$ per cubic centimeter~\cite{monoelectron, ebeam}.

The electron beam  density relavant is lower by a factor of $10^{4}$-$10^{5}$ than $10^{20}$ per cubic centimeter  and the electron beam energy is preferred to be comparable to a few hundred keV. 
The electron beam  does not need to be collimated either,  readily from the current beam technologies and might be generated in a much cheaper way or in a much smaller scale. 
The electron beam might be generated by various types of interaction between a  laser and any matter, including metals or some other available technologies.

A few examples for physical parameters in the THz generation are discussed below, wherein the electron beam parameter and the frequency of the microwave are illustrated. 
The first example has  a 100 keV electron beam with 300 GHz microwave. 
The critical density is  given as  $n_{c} = 4 \times 10^{15} $ per cubic centimeter, wherein the emitted light is 0.9 THz from Eq.~(\ref{eq:down1}).   
The second example has a 500 keV electron beam with  100 GHz microwave.
  The critical density is given as  $n_{c} = 9.6 \times 10^{14} $ per cubic centimeter, wherein the emitted light is 1.6 THz from Eq.~(\ref{eq:down1}).  
The third example has  a 1 MeV electron beam with 25 GHz microwave. 
  The critical density is  given as  $n_{c} = 0.45 \times 10^{14} $ per cubic centimeter, wherein the emitted light is 1 THz from Eq.~(\ref{eq:down1}).  
The fourth example has  a 2 MeV electron beam with 50 GHz microwave.
  The critical density is given as  $n_{c} = 3 \times 10^{14} $ per cubic centimeter, wherein the emitted light is 5 THz from Eq.~(\ref{eq:down1}).   
The last example has   a 2 MeV electron beam with 10 GHz microwave.  
The critical density is given as  $n_{c} = 1.2 \times 10^{13} $ per cubic centimeter, wherein the emitted light is 1 THz from Eq.~(\ref{eq:down1}).

As shown in the third, fourth and last examples, if the electron beam energy is as high as 1 MeV,  the frequency of microwave can be as low as or lower than 10 GHz. The gyrotron for such microwaves requires a magnetic field comparable to 1 Tesla instead of 10 Tesla, and the appropriate gyroron is   mobile and cheap.

The efficiency ratio can be still estimated based on Eq.~(\ref{eq:eff}). 
However, in the present scheme of the total reflection,  $n_{12} \cong 1000.0 $ and $\eta \cong 1$.  So that $E \cong 0.001 \times I_6$.  This is a bit lower than the scheme based on the plasma wave scattering. Furthermore, 
the electron beam on the present scheme needs to be generated by the laser-metal interaction, wherein the electron beam might not be directional, resulting that the efficiency ratio will be further lower. 
The current scheme is only possible to be operated on a pulsed duration. 

However, two key outstanding advantages of the current scheme is the peak intensity and the coherence. 
Due to the Doppler's effect, the peak intensity of the reflected THz light can be $4 \gamma_0^2 $ times of  that of the microwave.  Under the current technologies,
the intensity of the high-frequency microwave is as high as $10^{7} $ watt per square centimeter~[3] ($I_6 = 10 $).   If $\gamma_0 = 5$, then, the peak intensity of the THz light can be as high as as $10^{9} $ watt per square centimeter, which is very high. 
Furthermore, it will be a coherent light if the microwave is coherent.  

One limitation of the electron beam generation from the blow-out  would be the space charge build-up due to the expelled electrons. 
However, the background metal layers will provide the extra electrons for the charge and the return current, mitigating the space-charge build-up. 
Eventually, as the blowing-up process builds up, the electron density will get higher and the beam energy will gets lower, for an example, creating the strong magnetic field and jets.
The key optimization to be performed is to maintain the electron drift velocity between 50 keV and a few MeV with high enough density for the reflection. 
This optimization raises interesting experimental and theoretical questions. 

\section{Summary and Discussion}
In the conventional THz light generation based on an electron beam, 
the main mechanism is to channel the visible light laser or the wiggle by the conventional magnets into a THz light (FEL). The most powerful THz light source,
including  the free electron laser (FEL), is based on this channeling.
 However, its success is limited by the fact that 1) the preferred electron beam density (energy) should be high and  2)  the output of the THz light is extremly low.
The difficulties associated with  1) and 2) in the conventional approaches 
needs  more discussions in order to understand the key advantages of the current scenario. 


As for why the electron beam energy should be high, the visible light laser has wave frequencies between 30 THz and 300 THz 
and, in order to convert this laser into a THz light by the down-shift,  
the relativisitic factor of an electron beam should be between a few and a few tens, resulting that the preferred electron beam energy is a few MeV. 
The wiggle by the conventional magnets has the wave length of a few centimeter. 
In order to convert this wiggle into a THz light by the up-shift, 
the relativisitic factor of an electron beam should be between a few and a few tens, resulting that the preferred electron beam energy is also a few MeV.

As for why  the output of the THz light is extremly low,
in the conventional FEL, the interaction between the electron beam and the laser (the wiggle from periodic magnets) is the Thomson scattering, which is very weak. 
The most prominent method to achieve  a stronger scattering than the Thomson scattering is to  engage the plasma collective wave inside an electron beam.  For an example, 
the backward Raman scattering (BRS), which has been successfully utilized in the visible-light laser compression~[2] and constantly observed in the inertial confinement fusion research, has been demonstrated to be much stronger than the individual electron scattering. 
Considering an electron plasma with density of $10^{16} $ particles per cubic centimeter and a visible-light laser with intensity of $10^{15}$ watt per square centimeter, the BRS can be $10^{11}$ times larger than the conventional Thomson scattering.  
However, in order for the collective wave to interact effectively with the laser, the electron density is preferred to be in the range between $10^{18}$ per cc  and $10^{20}$ per cc   in order that   the plasmon frequency should be  comparable to the laser frequency. This electron density with a few MeV energy is expensive to create if not impossible. 

In this paper, a high frequency microwave is utilized instead of the conventional magnets or the visible-light laser.  
The preferred relativistic factor for up-shifting the microwave into the THz light  is  between 1 and 2, resulting that  the preferred electron beam energy is a few hundred keV instead of a few MeV. 
Another key advantages of the microwave is that 
the preferred electron density is between $10^{12}$ per cc and $10^{14}$ per cc (instead of between  $10^{18}$ per cc and  $10^{20}$ per cc)  in order for the collective plasma  wave frequency to be comparable to the microwave (visible light laser).  
The electron beam with such density and energy of a few hundred keV can be generated in much smaller machines such as a cathode and vitual cathode instead of an accelerator. Even if an electron beam is generated by an accelerator, the preferred quality or size of the accelerator is much more moderate or smaller.  
Due to this moderate requirement for the electron beam,    the size of the THz light source is compact and mobile.
The strong interaction of the plasma collective wave renders the THz light intensity  as high as or higher than the microwave. 

\bibliography{tera2}

\end{document}